\documentclass[aps,prl,superscriptaddress,twocolumn,floatfix]{revtex4-2}
\usepackage{graphicx,color}
\usepackage{dcolumn}
\usepackage{bm}
\usepackage{ulem}
\usepackage{latexsym}
\usepackage{amssymb}
\usepackage{amsfonts}
\usepackage{amsmath}
\usepackage{hyperref}
\usepackage{verbatim}
\usepackage{multirow}
\begin{document}
\preprint{APS/123-QED}

\title{Probing complex stacking in a layered material via electron-nuclear quadrupolar coupling}
\author{Li Cheng}
\affiliation{Shenzhen Geim Graphene Center (SGC), Tsinghua-Berkeley Shenzhen Institute (TBSI) and Tsinghua Shenzhen International Graduate School, Tsinghua University, Shenzhen 518055, China}

\author{Linpeng Nie}
\affiliation{Hefei National Research center for Physical Sciences at the Microscale, University of Science and Technology of China (USTC), Hefei 230026, China}

\author{Xuanyu Long}
\affiliation{Institute for Advanced Study, Tsinghua University, Beijing 100084, China}

\author{Li Liang}
\affiliation{Technology and Engineering Center for Space Utilization, Chinese Academy of Sciences, Beijing 100094, China}

\author{Dan Zhao}
\affiliation{Hefei National Research center for Physical Sciences at the Microscale, University of Science and Technology of China (USTC), Hefei 230026, China}

\author{Jian Li}
\affiliation{Hefei National Research center for Physical Sciences at the Microscale, University of Science and Technology of China (USTC), Hefei 230026, China}

\author{Zheng Liu}
\email{zheng-liu@tsinghua.edu.cn}
\affiliation{Institute for Advanced Study, Tsinghua University, Beijing 100084, China}

\author{Tao Wu}
\email{wutao@ustc.edu.cn}
\affiliation{Hefei National Research center for Physical Sciences at the Microscale, University of Science and Technology of China (USTC), Hefei 230026, China}
\affiliation{CAS Key Laboratory of Strongly-coupled Quantum Matter Physics, Department of Physics, University of Science and Technology of China, Hefei 230026, China}

\author{Xianhui Chen}
\affiliation{Hefei National Research center for Physical Sciences at the Microscale, University of Science and Technology of China (USTC), Hefei 230026, China}
\affiliation{CAS Key Laboratory of Strongly-coupled Quantum Matter Physics, Department of Physics, University of Science and Technology of China, Hefei 230026, China}

\author{Xiaolong Zou}
\email{xlzou@sz.tsinghua.edu.cn}
\affiliation{Shenzhen Geim Graphene Center (SGC), Tsinghua-Berkeley Shenzhen Institute (TBSI) and Tsinghua Shenzhen International Graduate School, Tsinghua University, Shenzhen 518055, China}
\date{\today}

\begin{abstract}
For layered materials, the interlayer stacking is a critical degree of freedom tuning electronic properties, while its microscopic characterization faces great challenges. The transition-metal dichalcogenide 1T-TaS$_2$ represents a novel example, in which the stacking pattern is not only enriched by the spontaneous occurrence of the intralayer charge density wave, but also recognized as a key to understand the nature of the low-temperature insulating phase. We exploit the $^{33}\rm{S}$ nuclei in a 1T-TaS$_2$ single crystal as sensitive probes of the local stacking pattern via quadrupolar coupling to the electron density distribution nearby, by combining nuclear magnetic resonance (NMR) measurements with the state-of-the-art first-principles electric-field gradient calculations. The applicability of our proposal is analyzed through temperature, magnetic-field, and angle dependent NMR spectra. Systematic simulations of a single 1T-TaS$_2$ layer, bilayers with different stacking patterns, and  typical stacking orders in three-dimensional (3D) structures unravel distinct NMR characteristics. Particularly, one 3D structure achieves a quantitative agreement with the experimental spectrum, which clearly rationalizes the coexistence of two types of interfacial environments. Our method may find general applications in the studies of layered materials.
\end{abstract}

\maketitle

\begin{figure*}
\centering
\includegraphics[width=15cm]{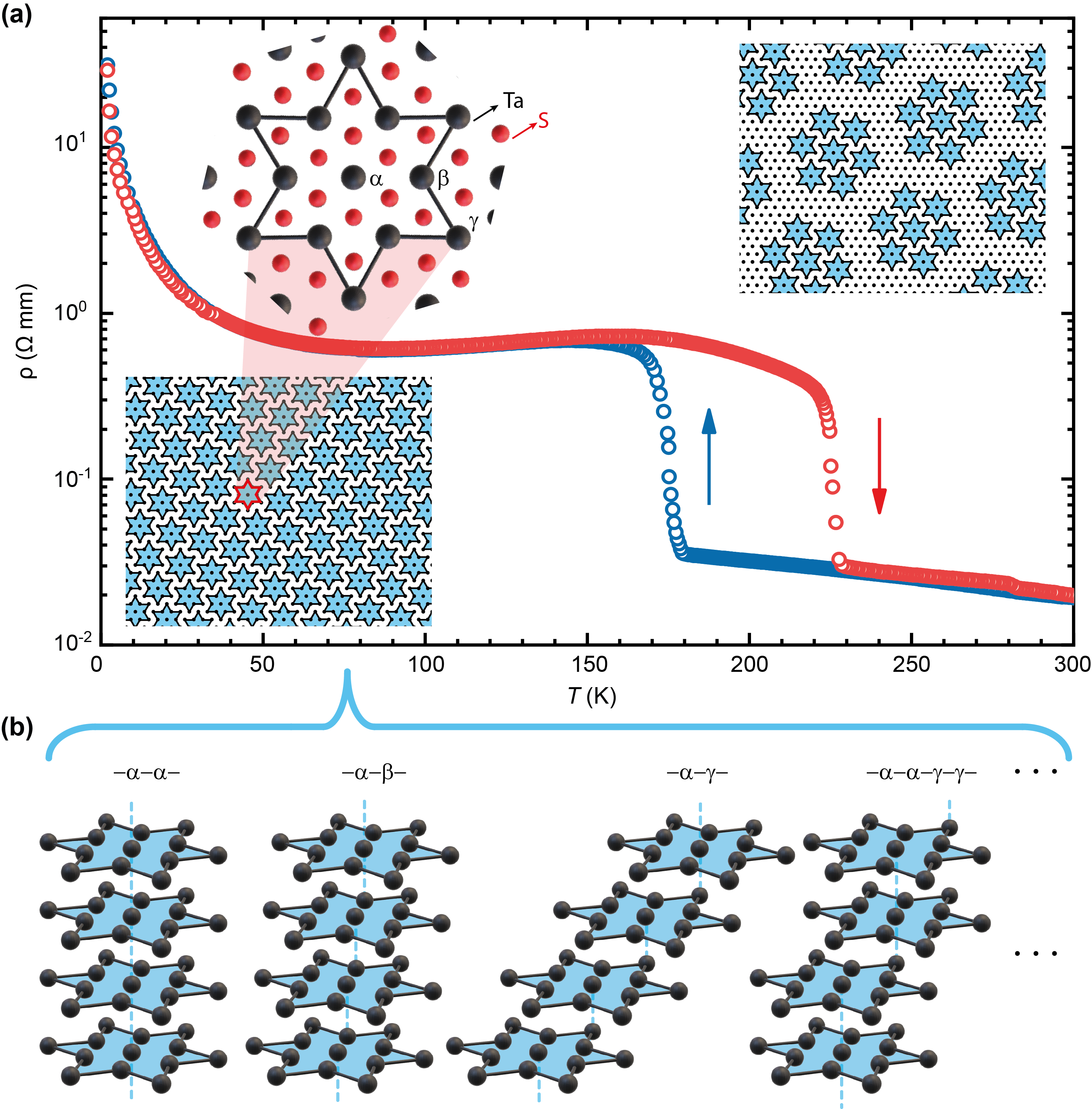}
\caption{Overview of CDW and inter-layer stacking in 1T-TaS$_2$. (a) The temperature-dependent resistivity measured in a heating (red) and cooling (blue) cycle. A first-order transition can be clearly seen around 200 K. The insets are the schematic SD patterns forming below the transition temperature and nearly commensurate CDW phase at higher temperature. Greek letters $\alpha$, $\beta$ and $\gamma$ label different Ta-sites within one SD. (b) The possible CDW stacking patterns of bulk 1T-TaS$_2$. Four representative structures are illustrated.}
\label{1}
\end{figure*}

The layered transition metal dichalcogenide (TMD) 1T-TaS$_2$ has been intensively studied for over four decades owing to its rich physics~\cite{APS_1975, PMB_1979, lee2017}. As temperature decreases, a series of charge density wave (CDW) transitions within the 2D layers occurs~\cite{Rossnagel_2011}. Associated with the formation of commensurate ($\sqrt{13}\times\sqrt{13}$)R13.9$^{\circ}$ Star-of-David (SD) patterns below around 200 K, the material enters an insulating state~[c.f. Fig. \ref{1}(a)]. Perturbing this insulating state can give rise to superconductivity, e.g. by pressurization~\cite{pressure2008} and chemical doping~\cite{Se-dope2013}.

If one views the long-period CDW ordering as a kind of  ``Moire'' pattern, the above scenario appears similar to twisted bilayer graphene at the magic angle~\cite{Caoyuan43,Caoyuan80,science.aav1910}, in particular by noticing that an isolated flat band around the Fermi level is created~\cite{lichengprb22}. A common theoretical practice is to explore the strongly-correlated physics within the flat band subspace~\cite{PRL14Nandini,PRL18spinon}, which is arguably even more exotic in 1T-TaS$_2$ due to the geometry frustration of the TMD triangular lattice~\cite{lee2017,2017high}.

Such an understanding faces challenges when the inter-layer coupling is also taken into account. By considering a bilayer stacking order, first-principles band calculations can indeed obtain an insulating state without invoking Mottness~\cite{Natphys_Geck_2015,2018stacking,prl2019lee}. Experimentally, signatures of bilayer stacking can be observed in X-Ray diffraction data~\cite{NC2020Zhang}. However, the weak intensity of the half-integer diffraction peaks and the broadening of the CDW diffraction peak suggest additional complexity and certain randomness~\cite{NC2020Zhang}, as noticed for a long time from other experimental methods~\cite{TaNQR1984,FUNG198047}. Recent scanning tunneling microscope measurements have identified different stacking patterns on the sample surfaces, showing varying dI/dV spectra
~\cite{NC2020Butler,LeePRL2021,YinYiPRB2022}. By shuffling the relative positions of the SD centers, the number of possible stacking patterns is huge [Fig. \ref{1}(b)]. So far, tracking the intriguing stacking pattern of bulk 1T-TaS$_2$ remains technically difficult.

The nuclear hyperfine levels are sensitive to the electronic environment nearby, based on which powerful experimental techniques, such as nuclear magnetic resonance (NMR),  nuclear quadrupole resonance (NQR) and Mossbauer spectroscopy, are developed. For our purpose, given that nuclei with nonzero electric-quadrupole moment are coupled to the electric-field gradient (EFG) generated by electrons, it can be used to probe delicate electron redistribution associated with different stacking patterns. Empirical analysis based on early $^{181}$Ta NMR~\cite{1986}, NQR~\cite{TaNQR1984}, as well as time differential perturbed angular correlation spectroscopy~\cite{181Ta1990} ruled out a simple stacking order, and led to proposals of several complex stacking patterns. This present work aims to promote the characterization ability of this type of techniques to a new level by combining with the state-of-the-art first-principles EFG calculations, which enables direct and sharpened confrontation between theory and
experiment.

Our synergistic study makes progress in two aspects: (i) the hyperfine level of $^{33}$S instead of $^{181}$Ta is detected; (ii) a systematic first-principles procedure is designed to interpret the NMR spectrum. Considering the S-Ta-S sandwiched structure of each 1T-TaS$_2$ layer and the larger number of inequivalent S-sites compared to Ta-sites, a switch from Ta to S as the local probe is expected to better resolve the stacking effects. More importantly, the light $^{33}$S nuclei can be much more reliably described within the standard density functional theory (DFT) framework than $^{181}$Ta. For the latter, an incomplete treatment of relativistic effects could lead to significant computational errors. The computational procedure as demonstrated here should find general applications in other 2D materials as well.

The high-quality single crystals of 1T-TaS$_2$ were grown by the standard chemical vapor transport method. Tantalum and sulfur power, being thoroughly mixed with 150 mg iodine as the transport agent, were sealed into an evacuated quartz tube. It should be noted that the powder of sulfur is enriched with isotope $^{33}\rm{S}$ for the purpose of NMR experiment on sulfur sites. The quartz tube was placed into a two-zone furnace with the temperature from 1303 K to 1173 K and then maintained for 10 days. After the crystal growth of 1T-TaS$_2$, the quartz tube is quickly quenched into ice water to avoid the formation of 2H-TaS$_2$ phase. Finally, the crystals were washed with ethanol to remove iodine on the surface. Resistance measurement was carried out with the standard four-probe method in a commercial Physical Property Measurement System (PPMS) from Quantum Design company. The NMR measurement was performed with a commercial NMR spectrometer from Thamway company and a 12 T magnet from Oxford Instruments company. The external magnetic field is calibrated by the $^{63}\rm{Cu}$ NMR signal from the NMR coil. All the NMR spectra were measured by a standard spin echo method and analyzed by a fast Fourier transform method.

Our calculations are performed by using VASP~\cite{kresse1996efficient,kresse1996}, and the built-in PAW data set~\cite{kresse1999ultrasoft}. The EFG computation module in VASP follows the method described in Ref.~\cite{EFGPAW}. A high plane wave energy cutoff (500 eV for all the EFG calculations) and a dense k-point mesh ($4\times4\times8$ for a simple stacking order with intralayer $\sqrt{13}\times\sqrt{13}$ CDW periodicity) are used to achieve high convergence of the electron density. The electronic self-consistent iterations are converged to 10$^{-8}$ eV precision of the total energy. An effective $U = 2.27$ eV as previously used for this material~\cite{qiao2017, lichengprr20} is introduced within the DFT+U framework~\cite{PRB2014Darancet} to correct the Coulomb interaction of Ta 5d electrons. The experimental in-plane lattice constants ($a_0$ = $b_0$ = 3.36 ${\AA}$)~\cite{explattice} are used to construct the simulation supercell.

\begin{figure*}
\centering
\includegraphics[width=15cm]{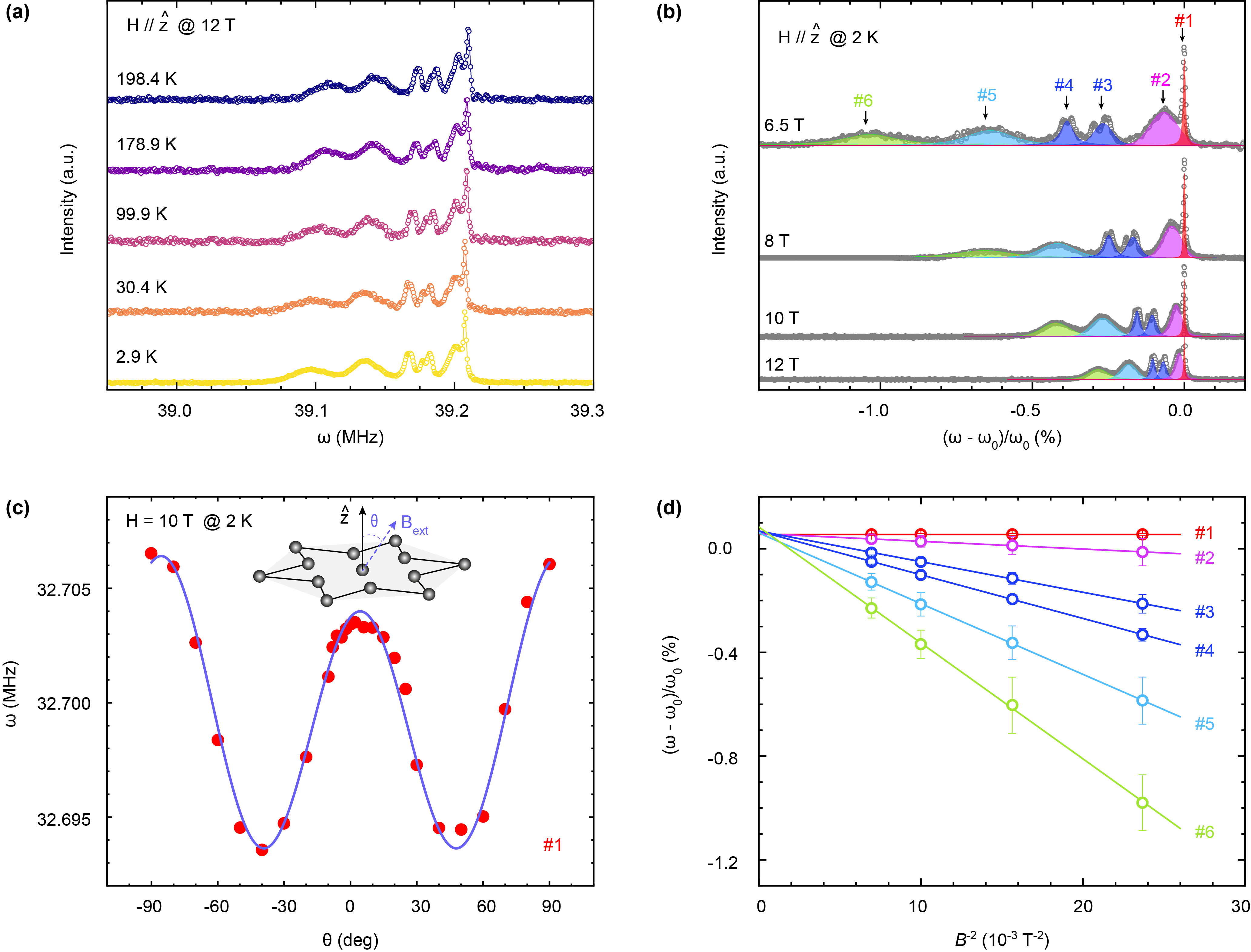}
\caption{Experimental analysis of the NMR spectrum. (a) Temperature-dependent $^{33}$S spectra with the external field (12 T) parallel to $\hat{z}$-axis. (b) Field-dependent $^{33}$S spectra at 2 K. The gray dots are experiment data, the colored shadows are Gaussian fitting curves. (c) The frequency variation of peak $\#$1 $^{33}\rm{S}$ at 2 K and 10 T (red dots), as a function of the angle between the external field and the $\hat z$-axis (inset). The 
curve is the result of Eq.(~\ref{eq:omegaq}) with $V_{ZZ}Q=1.9$ MHz and $\eta=0$. (d) The field-dependent frequency shift of the peaks numbered in (b). The error bars are defined as the full width of half maximum of the Gaussian fitting curves.}
\label{2}
\end{figure*}

Figures~\ref{2} shows the single-crystal NMR data of $^{33}\rm{S}$, which has a nuclear spin $I=\frac{3}{2}$, a gyromagnetic ratio $\gamma = 3.2655$ MHz/T and a quadrupole moment $Q = -67.8$ millibarn~\cite{Q2008}. The measured spin-echo resonance frequency ($\omega$) corresponds to the central $-\frac{1}{2}\leftrightarrow \frac{1}{2}$ magnetic quantum number transition. 
Within the commensurate CDW phase, the spectrum is nearly temperature independent [Fig. \ref{2}(a)]. The multi-peak structure of the spectrum indicates a large number of inequivalent S-sites, as a natural consequence of intralayer CDW and interlayer stacking. 

With Zeeman splitting $\omega_0=\gamma B$ as the largest energy scale, the quadrupolar coupling $H_Q = \frac{V_{ZZ}Q}{4I(2I-1)}\left[3\hat{I}^2_Z-I(I+1)+\eta(\hat{I}^2_{X}+\hat{I}^2_{Y})\right]$ can be treated as a weak perturbation (See e.g. Chpt. 10 in \cite{NMR_CP}), where $I_{X,Y,Z}$ are the angular momentum operators of the nuclear spin. $H_Q$ contains two EFG parameters $V_{ZZ}$ and $\eta$, corresponding to the largest component and the $XY$ asymmetry parameter of the EFG tensor, which are site and stacking dependent.

A special property of the $-\frac{1}{2}\leftrightarrow \frac{1}{2}$ transition~\cite{QIchapter} is that the first-order correction of the quadupolar couping vanishes, and the second-order correction gives:
\begin{equation}\label{eq:omegaq}
\delta \omega_{Q}=-\frac{1}{8}\frac{V_{zz}^2Q^2}{\omega_0} [A(\phi, \eta) \cos ^{4} \theta+B(\phi, \eta) \cos ^{2} \theta+C(\phi, \eta)],
\end{equation}
with
\begin{equation}\label{eq:ABC}
\begin{aligned}
&A(\phi, \eta)=-\frac{27}{8}+\frac{9}{4} \eta \cos 2 \phi-\frac{3}{8}(\eta \cos 2 \phi)^{2} \\
&B(\phi, \eta)=\frac{30}{8}-\frac{1}{2} \eta^{2}-2 \eta \cos 2 \phi+\frac{3}{4}(\eta \cos 2 \phi)^{2} \\
&C(\phi, \eta)=-\frac{3}{8}+\frac{1}{3} \eta^{2}-\frac{1}{4} \eta \cos 2 \phi-\frac{3}{8}(\eta \cos 2 \phi)^{2}.
\end{aligned}
\end{equation}
$\theta$ and $\phi$ are the polar angles of the external magnetic field with respect to the principle axes of the EFG tensor, which are also site and stacking dependent. We will refer to $\delta \omega_Q$ as the quadrupolar shift (QS). 

According to Eq.(\ref{eq:omegaq}), when the direction of the external field is fixed, $\delta \omega_{Q}/\omega_0\propto \omega_0^{-2}\propto B^{-2}$. Figure~\ref{2}(b) shows the frequency shift relative to the Zeeman splitting measured under four different $\hat z$-direction external fields. We fit each spectrum with six Gaussian envelopes [Fig.~\ref{2}(b)], and the fitted peak positions are plotted in Fig.~\ref{2}(d) against $B^{-2}$; a good linear scaling is found. Since the QS is suppressed as $B$ increases, the intercept at $B^{-2}=0$ reflects frequency shift of other origins, e.g. magnetic shift induced by electron spin polarization. 
Crucially, the intercept converges to a very narrow range, compared to the spectrum width, presenting as a site-independent constant ($\sim 5\times10^{-4}\omega_0$). In our calculations below, this non-QS effect is neglected. 

As a further justification of the predominant role of QS, we fixed the strength of the external field and measured the angle dependence of the spectrum. Figure~\ref{2}(c) shows the frequency of peak $\#1$ as a function of $\theta$, and the experimental data can be nicely fitted by Eq.(\ref{eq:omegaq}) with $V_{ZZ}Q=1.9$ MHz and $\eta=0$.

We now turn to  first-principles simulations of the NMR spectrum. Based on Eq.(\ref{eq:omegaq}), once the full EFG tensor associated with a S-site is obtained, all the parameters in Eq.(\ref{eq:omegaq}) are fully decided; the QS can then be quantitatively calculated. Essentially, the EFG tensor is defined by the second derivatives of the electrostatic potential. To derive it, the input is actually nothing but the ground-state electron density, which well fits the capabilities of DFT. Thanks to the long-time efforts to improve the density precision around nuclei, e.g. by means of linearized augmented plane wave basis~\cite{EFGLAPW} or projector augmented wave (PAW) potential~\cite{EFGPAW}, nowadays standard output of the EFG tensor has been implemented in many widely-used first-principles packages~\cite{VASPchapter,abinit,wien2k}.

\begin{figure}
\centering
\includegraphics[width=8cm]{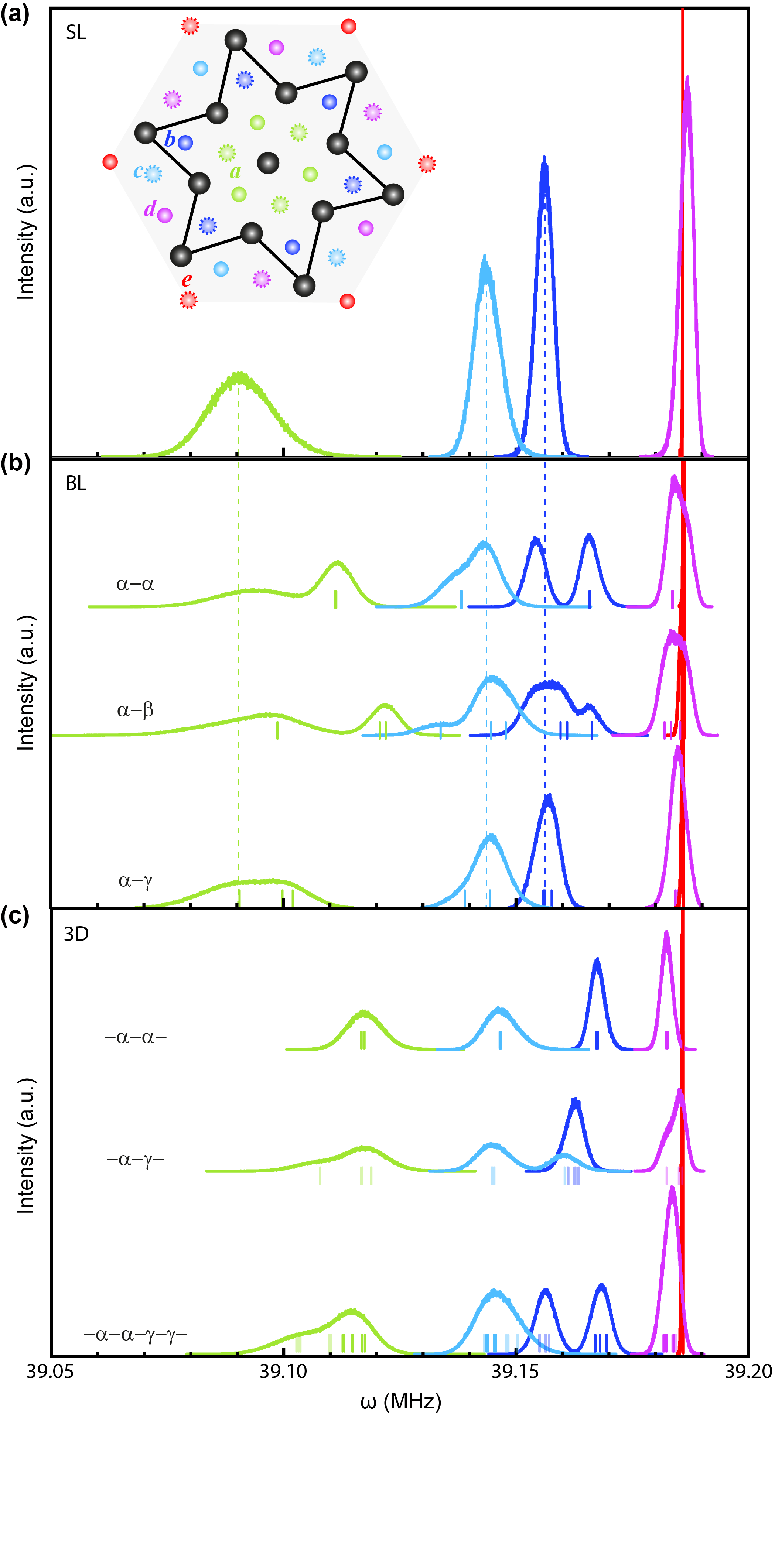}
\caption{Computational analysis of the NMR spectrum. (a), (b) and (c) are calculated NMR spectra of a SL, BLs and 3D periodic structures with different stacking patterns. The external field is considered to be applied along the $\hat{z}$-aixs with a fixed strength 12 T. Colors represent the S-sites as labeled in the inset of (a), where the solid and hollow circles further differentiate the upper and lower S layers. Dashed lines indicate the peak positions in SL spectrum. The short vertical lines in (b) and (c) represent the resonant frequencies of S sites at interfaces. For the BLs, the outward S sites facing the vacuum are not explicitly marked by short vertical lines, which largely retain the SL feature. For the 3D structures, the deep and shallow colors of the short vertical lines are used to denote S sites at the $\alpha$-$\alpha$ and $\alpha$-$\gamma$ interfaces, respectively.} 
\label{3}
\end{figure}

\begin{table}[!ht]
\caption{Calculated EFG parameters of a SL.}
\begin{center}
\scalebox{0.8}{
\setlength\extrarowheight{8pt}
    \begin{tabular}{cccccc}
    \hline
         {Site $\#$}& $a$ & $b$ & $c$ & $d$ & $e$  \\
    \hline
        {$V_{ZZ}Q$ (MHz)}& 5.422& 6.496& 5.855& 3.211& 1.732 \\
        {$\eta$}& 0.778& 0.677& 0.411& 0.802& 0.000 \\
        {$\theta$($^{\circ}$)}& 145.927& 38.080& 46.356& 79.224& 0.000 \\
        {$\phi$($^{\circ}$)}& 89.726& 174.681& 11.911& 48.195& 90.000 \\
    \hline
    \end{tabular}}
    \label{tab1}
\end{center}
\end{table}

We start from calculating a single 1T-TaS$_2$ layer (SL). The external field is considered to be applied along the $\hat{z}$-aixs with a fixed strength 12 T. To simulate the commensurate CDW order, a $\sqrt{13}\times\sqrt{13}$ supercell containing 13 Ta atoms is employed. A 18 ${\AA}$ vacuum layer normal to the 2D plane is included to avoid interlayer interactions. All of the atoms are relaxed until the atomic forces are less than 0.02 eV/${\AA}$. The intralayer SD pattern forms automatically during structural relaxation. As labeled in the inset of Fig.~\ref{3}(a), there are five inequivalent S-sites within one SD according to the S$_6$ point group, giving rise to five different EFG tensors (Tab.~\ref{tab1}). Substituting the calculated EFG parameters into Eq.(\ref{eq:omegaq}) results in five discrete resonance frequencies. To better simulate the experimental spectrum, we assume that arising from either experimental or computational sources, $V_{ZZ}$ and $\eta$ are subject to a $2\%$ relative uncertainty, while $\theta$ and $\phi$ are subject to a 2 degree absolute uncertainty, which effectively broaden the five discrete resonance frequencies into continuum envelopes. Numerically, these uncertainties are added as Gaussian noises to the calculated EFG parameters, and make the histogram plot. The integrated area spanned by each envelop is proportional to the corresponding site population. We note that the choice of the broadening width is arbitrary, but the position of the central peak is unchanged. For the multilayer calculations, we employ exactly the same frequency broadening treatment without further fine tuning.

The simulated SL spectrum [Fig.~\ref{3}(a)] already captures some novel features of the experimental data. The strong and sharp peak $\#$1 labeled in Fig.~\ref{2}(b) can be unambiguously attributed to site $e$, the only high-symmetry S-site. The associated C$_3$ site-symmetry group requires that $V_{ZZ}$ is exactly perpendicular to the 2D plane ($\theta$=0), and there is no $XY$ asymmetry ($\eta$=0), corroborating Fig.~\ref{2}(c). It is also worth mentioning that the calculated value of $V_{ZZ}Q$ (1.732 MHz) well agrees with the experimental value (1.9 MHz) fitted from Fig.~\ref{2}(c). Since $\delta\omega_Q$ vanishes at $\theta$=0 and $\eta$=0, the distance to peak $\#$1 is a direct measure of QS. Site $a$ turns out to experience the largest QS with a negative sign, consisting of the broad distribution at the low-frequency side. In contrast, site $d$ experiences a small positive QS, lying contiguously right to the site $e$ peak,  in together giving rise to the high-frequency resonance. The site $b$ and $c$ peaks locate approximately in the middle of the spectrum. According to Tab.~\ref{tab1}, there is no single parameter distinguishing the low symmetry sites, for which the resonance frequencies can hardly be estimated without first-principles calculations. 

The inter-layer coupling is expected to further shift and split these SL peaks. To see these effects, we construct three 2D bilayer (BL) structures and three 3D periodic stacking structures. The interlayer spacing is determined by energy minimization, and the van der Waals interaction is added to the DFT calculations within the Tkatscenko-Scheffler scheme \cite{TSvdw2009}. By using the SD center [Ta $\alpha$-site; see the Greek letters labeled in the inset of Fig.~\ref{1}(a)] of one layer as the reference, the adjacent layer is aligned by placing one of the Ta sites of the SD on top. We note that it is unrealistic to perform exhaustive testing on all possible stacking patterns. Instead, we focus on the representative ones discussed in literatures. The $\alpha$-$\alpha$, $\alpha$-$\beta$ and $\alpha$-$\gamma$ BLs in our notation correspond to the AA, AB and AC stacking identified on the cleaved surfaces of 1T-TaS$_2$ samples by STM~\cite{YinYiPRB2022}. The -$\alpha$-$\alpha$-, -$\alpha$-$\gamma$- and -$\alpha$-$\alpha$-$\gamma$-$\gamma$- 3D structures correspond to the A, L and AL stacking orders theoretically investigated in Ref.~\cite{prl2019lee}.

The calculation results show that peak $\#$1 is largely free from the stacking effect. Although the C$_3$ site-symmetry group does not necessarily exist under stacking, the modified EFG parameters only changes $\delta \omega_Q$ marginally. For the BLs, the outward S-sites facing the vacuum are found to largely retain their resonance frequencies as in the SL [See the peaks around the dashed lines in Fig.~\ref{3}(b)]. This indicates that the stacking effect on EFG is short ranged. The inward S-sites at the interface show clearly modified QS [peaks marked by short vertical lines in Fig.~\ref{3}(b)]. 
An important improvement for all the BLs is that the site $d$ peak is pushed to the low-frequency side, making the sharpest site $e$ peak the highest-frequency one, in agreement with the experiment. 

\begin{figure}
\centering
\includegraphics[width=8cm]{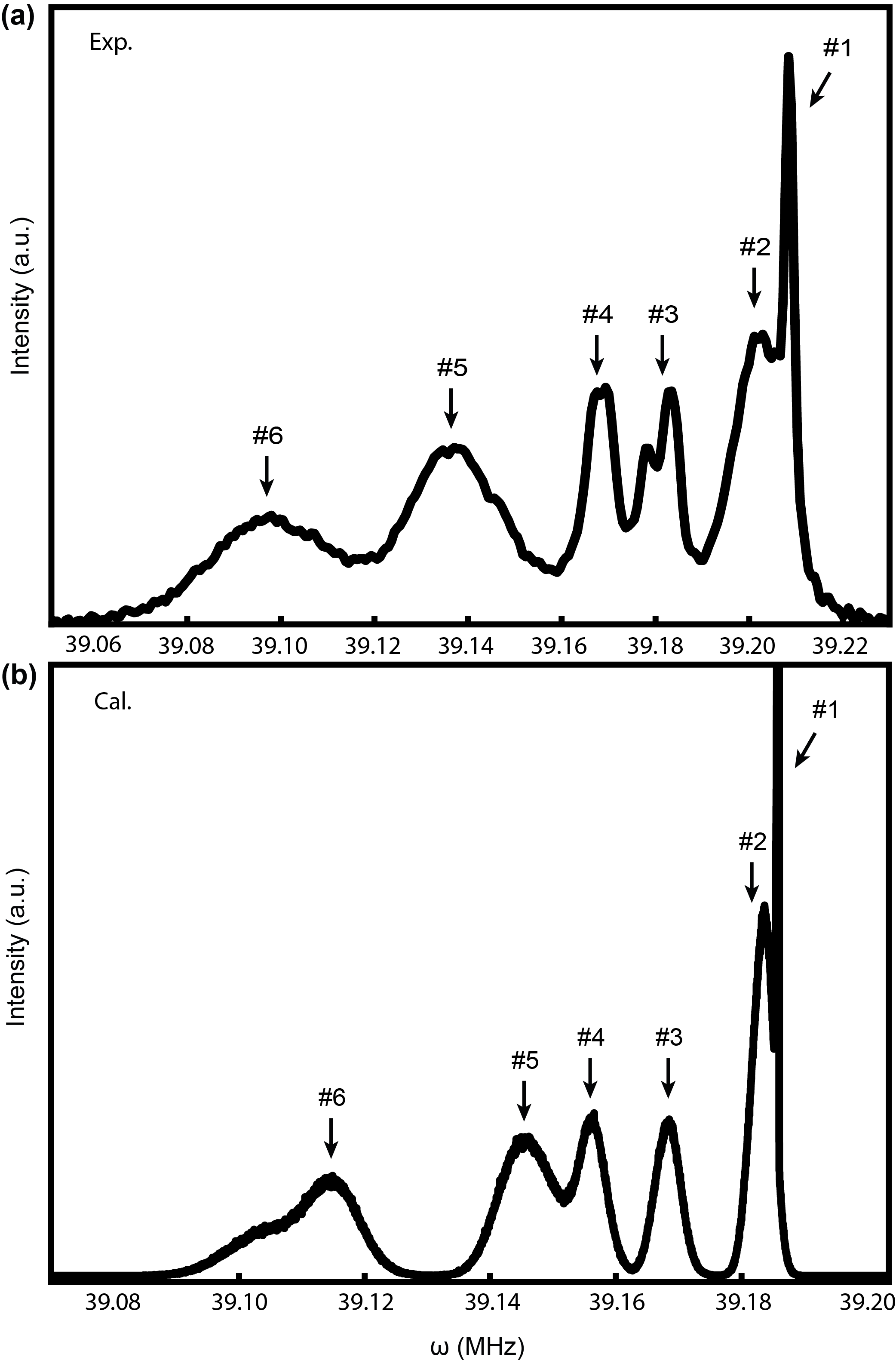}
\caption{Direct comparison between the experimental NMR spectrum and the first-principles simulation. (a) The full NMR spectrum of $^{33}\rm{S}$ at 2 K with external magnetic field $B$ = 12 T normal to the TaS$_2$ 2D layers. (b) The calculated spectrum of a -$\alpha$-$\alpha$-$\gamma$-$\gamma$- 3D periodic structure. Note that the frequency axes of the two figures are not identical, and the calculation conisiders the QS only.}
\label{4}
\end{figure}

Among the simulated stacking patterns, the spectrum of the -$\alpha$-$\alpha$-$\gamma$-$\gamma$- structure resemble the experiment best. Such a stacking pattern incorporates the
effects of two types of interfacial environments [$\alpha$-$\alpha$ and $\alpha$-$\gamma$, with their resonance frequencies differentiated by the brightness of the short vertical lines in Fig.~\ref{3}(c)]. We note that the energy difference between -$\alpha$-$\gamma$- and -$\alpha$-$\alpha$-$\gamma$-$\gamma$- stackings were found to be tiny in previous first-principles calculation~\cite{prl2019lee}, whereas their NMR spectra differ more significantly.

A key difference between the -$\alpha$-$\alpha$- and -$\alpha$-$\alpha$-$\gamma$-$\gamma$- spectra is the number of peaks. The -$\alpha$-$\alpha$- stacking inherits the full S$_6$ point group of the SL, so by symmetry there should be no additional peak splitting. In contrast, the only remaining symmetry operation for the -$\alpha$-$\alpha$-$\gamma$-$\gamma$- stacking is an inversion center in the middle of two Ta layers. The most remarkable additional splitting arises from site $b$ [colored in blue in Fig.~\ref{3}], which is also evident in experimental spectra [Peaks $\#$3 and $\#$4 in Fig.~\ref{2}(b)]. Assigning peaks $\#$3 and $\#$4 to the same site is strongly backed by inspecting their intensity, which is proportional to the site population. It is straightforward to count [c.f. inset of Fig.~\ref{3}(a)] that the numbers of sites-$a$, $b$, $c$, $d$ associated with one SD are equal, while that of site-$e$ is one third of the formers. The integrated area of each colored envelope in Fig.~\ref{2}(b) exactly coincides with this ratio. In particular, peaks $\#$3 and $\#$4 in together carries the expected intensity. Similar to the -$\alpha$-$\alpha$- stacking, this characteristic site $b$ splitting is absent in the -$\alpha$-$\gamma$- 3D structure. It suggests that the coexistence of two types of interfacial environments ($\alpha$-$\alpha$ and $\alpha$-$\gamma$) should be considered as an essential ingredient to interpret the experimental data. 

We  should also mention that since NMR is a local probe, the periodicity assumed in the -$\alpha$-$\alpha$-$\gamma$-$\gamma$- structure is not a necessary condition to reproduce the experimental data. For example, a simple superposition of the -$\alpha$-$\alpha$- and -$\alpha$-$\gamma$- spectra shown in Fig.~\ref{3}(c) can largely reproduce the experimental results as well, but again, the key is the coexistence of two types of interfacial environments. Therefore, our study does not exclude stacking disorders, stacking faults or rotations of the $\alpha$-$\gamma$ shift direction. The fine structure around each main peaks of the experimental spectrum indeed reflects the existence of additional complexity.

In summary, Figure~\ref{4} shows a direct comparison between the experimental and computational NMR spectra based on the -$\alpha$-$\alpha$-$\gamma$-$\gamma$- 3D structure. Since the calculation neglects any non-QS effect, the calculated spectrum width is slightly narrower. Nevertheless, considering that there is no fitting parameters in the calculation, we find that such an agreement is readily persuasive. This sharpened confrontation provides strong evidence of coexistence of the $\alpha$-$\alpha$ and $\alpha$-$\gamma$ interfaces in bulk 1T-TaS$_2$, leaving little room for a simple stacking pattern. It is worth mentioning that $^{33}\rm{S}$ can also be used to probe the nature of the insulating state via the nuclear spin relaxation behavior. The NMR data are currently under analysis and will be published elsewhere.



This work is supported by the National Natural Science Foundation of China (Grants No. 11974197, 51920105002, 12034004, 12161160316), the National Key R\&D Program of the MOST of China (Grants No. 2022YFA1602601), Guangdong Innovative and Entrepreneurial Research Team Program (No. 2017ZT07C341) and Tsinghua University Initiative Scientific Research Program.

L.C. and L.N. contributed equally to this work.

\bibliography{ref} 
\end{document}